\documentclass[preprint,showpacs,preprintnumbers,amsmath,amssymb, showkeys]{revtex4}

\usepackage{array}
\usepackage{booktabs}
\usepackage{tabu}
\usepackage{dcolumn}
\usepackage{amsmath}
\usepackage{amsfonts}
\usepackage{amssymb}
\usepackage{graphicx}
\usepackage{subfigure}
\usepackage{graphicx}
\usepackage{dcolumn}
\usepackage{bm}

\def\be{\begin{equation}}
\def\ee{\end{equation}}
\def\bea{\begin{eqnarray}}
\def\eea{\end{eqnarray}}
\def\f{\frac}
\def\n{\nonumber}
\def\l{\label}
\def\p{\phi}
\def\o{\over}
\def\R{\rho}
\def\pa{\partial}
\def\om{\omega}
\def\na{\nabla}
\def\P{\Phi}

\begin{document}

\title{Hamilton-Jacobi formalism to warm anisotropic inflation }

\author{M. Tirandari}
 \email{mehraneh.tirandari@gmail.com}
 \affiliation{
 Department of Physics, Sharif University of  Technology,Tehran 11155-9161, Iran}
 \author{A. Mohammadi}
 \email{abolhassanm@gmail.com}
\affiliation{
Department of Physics, University of Kurdistan, Pasdaran St., P.O. Box 66177-15175, Sanandaj, Iran}
\date{\today}

\def\be{\begin{equation}}
\def\ee{\end{equation}}
\def\bea{\begin{eqnarray}}
\def\eea{\end{eqnarray}}
\def\f{\frac}
\def\n{\nonumber}
\def\l{\label}
\def\p{\phi}
\def\o{\over}
\def\R{\rho}
\def\pa{\partial}
\def\om{\omega}
\def\na{\nabla}
\def\P{\Phi}

\begin{abstract}
Warm inflation in the framework of locally rotationally symmetric Bianchi type I universe model in the Hamilton-Jacobi formalism is being considered. Matter-Radiation fluid in Warm inflation, endowed with and without a viscous pressure. For the tow different cases dynamical equations are obtained under slow-roll approximation and in the strong dissipative regime. Special cases of dissipation and bulk viscous pressure coefficients are considered. Warm anisotropic inflationary model is compatible with observational data. But for Warm anisotropic inflation with viscous pressure in the first case where dissipation and bulk viscous pressure coefficients are considered to be constant, perturbation parameters is not compatible with observational data. In the case where dissipation and bulk viscous pressure coefficients are considered to be variable, the perturbation parameters is compatible with Planck data.
\end{abstract}
\maketitle

\section{Introduction}
\label{S:1}
The inflationary universe scenario is of particular importance to cosmology since it may naturally solve many of the Big Bang model problems. The idea of inflation is that an early phase of accelerated expantion can serve to stretch out any primordial inhomogeneities and create an almost flat universe. In standard inflationary models, acceleration of universe is driven by a scalar field with a specific scalar potential \cite{Guth}-\cite{Starobinsky}. This model has two regimes, the slow-roll and reheating period. The latter is a way to attach the observed universe to the end of inflationary epoch. There are two other way to do this: preheating and warm inflation. \\
warm inflation is an alternative mechanism to have inflation without reheating period \cite{Berera}-\cite{Hall}. In this kind of model inflaton field interact with other fields. So, dissipative effects become strong due to the production of thermalize particles. This effect can be represented by adding a friction term to the equation of motion for the scalar field. Also, radiation is not completely diluted away and the friction term in the conservation equation is accounting for transfer of energy between the inflaton and this radiation. However accelerated expantion only occurs for sub-dominant radiation, $ \rho_{r}\ll\rho_{\phi} $. Moreover, thermal fluctuations in the radiation are transfered to the inflaton and become the main source of primordial fluctuations \cite{Berera}. In this form the essential condition for warm inflation to occur is that the radiation temperature is $ T>H $, since thermal fluctuations is propertional to $ T $ and the quantum onse is proportional to $ H $.\\
particle production alters the dynamics of fluid. It  modifies hydrodynamic pressure to a more general expression $ p=(\gamma-1)\rho $ with $ \rho $ the energy density of the matter-radiation fluid and $ \gamma $ the adiabatic index, which is bounded by $ 1\leqslant\gamma\leqslant 2 $. Moreover, it produces a non-equilibrium, viscous, pressure $ \Pi $ via two different mechanisms: (i) inter-particle interaction. (ii) the decay of particle within the fluid \cite{cid}- \cite{Mimoso}. \\
Several aspects of warm inflationary universe models have been studied in liturature \cite{cid2}-\cite{Bastero-Gil2}. The main aim of this work is to investigate warm inflationary models in an anisotropic BI space time in the so called Hamilton-Jacobi formalism \cite{Salopek}. In this approach, the scalar field is considered to be time variable. It allows one to consider $ H(\phi) $, rather than $ V(\phi) $, as a fundamental quantity to be specified \cite{LiddleLyth}. This formalism has been considered for warm inflation in the FRW universe by \cite{Sayar} and \cite{Akhtari}. 
Our work is organized as follows. In the next section we introduce BI metric. In section \ref{S:3}, at first we describe a brief reviw of the warm anisotropic inflation in the Hamilton-Jacobi formalism, and than we apply it to a specific example. We calculate the tensor-to-scalar ratio, and the spectral index and its running, and we compare them with the Planck data. In section \ref{S:4} we describe  the warm anisotropic inflation with viscous pressure in the Hamilton-Jacobi formalism. Then we take an example and we investigate purturbation parameters and compare them with Planck data. In section \ref{conclusion} we make conclusion of the work. 
 

\section{Anisotropic metric}
\label{S:2}

We consider a BI univers filled with a self-interacting scalar field $ \phi $ and an imperfect fluid. Anisotropic BI univers is considered to be as follows

 \begin{equation}
\label{BImetric}
ds^{2}=-dt^{2}+a^{2}(t)dx^{2}+b^{2}(t)(dy^{2}+dz^{2}),
\end{equation}
where $ a(t) $ is the scale factor to measure expansion along $ x $ direction  and $ b(t) $ is scale factor to measure expansion along $ y $ and $ z $ direction. The average Hubble expantion is defined as
\begin{equation}
\label{Habble}
H_{t}=\frac{1}{3}(H_{1}+2H_{2}), \  \ H_{1}=\frac{\dot{a}(t)}{a(t)}, \  \  H_{2}=H_{3}=\frac{\dot{b}(t)}{b(t)},
\end{equation}
 A linear relationship, $ a=b^{\lambda} $, ($ H_{1}=\lambda H_{2} $) reduces the above equation to 
\begin{equation}
\label{Habble2}
H=\frac{1}{3}(\lambda+2)H_{2}
\end{equation}
where $ \lambda $ ($ \lambda> 1  $) is a constant and represents a deviation from the isotropy along $ x $ direction. $ \lambda= 1  $ corresponde to FRW space time. \\

\section{Warm anisotropic inflationary model }
\label{S:3}

\subsection{Generality}
Enegy density and pressure of the inflaton are as follows
\begin{equation}
\label{density11}
\rho_{\phi}=\frac{1}{2}\dot{\phi}^{2}+U(\phi), \  \ p_{\phi}=\frac{1}{2}\dot{\phi}^{2}-U(\phi),
\end{equation}
If the relativistic particles thermalize fast enough, we can model their contribution as that of radiation:
\begin{equation}
\label{radiationdensity}
\rho\simeq\frac{\pi^{2}}{30}g_{*}T^{4}=C_{\gamma}T^{4},
\end{equation} 
where T is the temperature of the thermal bath, and  $ g_{*} $ the effective number of light degrees of freedom.\\
The dynamics of anisotropic warm inflation without viscous pressure is given by the corresponding Friedmann equation as follows
\begin{equation}
\label{friedman}
H_{2}^{2}=\frac{1}{1+2\lambda}(\rho_{\phi}+\rho),
\end{equation}
The inflaton decays with rate $ \Gamma $ into the imperfect fluid. The conservation equations for the inflaton is 
\begin{equation}
\label{conservation3}
\ddot{\phi}+((\lambda+2)H_{2}+\Gamma)\dot{\phi}+U^{'}(\phi)=0,
\end{equation}
where the prime denotes derivative with respect to $ \phi $. An upper dot indicates temporal derivative. In general, the dissipation coefficient $ \Gamma $ depends on $ \phi $ or $ T $ or both, and as a consequence of second law of thermodynamics $ \Gamma>0 $. It has been computed in \cite{Moss}, using the near-equilibrium approximation. Also, the radiation field is governed by the transfer relation
\begin{equation}
\label{conservation2}
\dot{\rho}+(\lambda+2)\frac{4}{3}H_{2} \rho=\Gamma\dot{\phi}^{2}.
\end{equation}
By considering quasi-stable decay of inflaton into imperfect fluid i.e $ \dot{\rho}\ll(\lambda+2)\frac{4}{3}H_{2} \rho$ and $ \dot{\rho}\ll\Gamma\dot{\phi}^{2} $, energy density of the radiation reads
\begin{equation}
\label{density4}
\rho=\frac{3\Gamma }{4(\lambda+2)H_{2}}\dot{\phi}^{2}=\frac{3 }{4}Q\dot{\phi}^{2}
\end{equation}
where we have introduced the dissipative ratio $ Q=\Gamma/((\lambda+2)H_{2}) $. This ratio is not necessarily constant and depending on the model it may increase or decrease during inflation which distinguishes between strong ($ Q\gg1 $) and weak dissipative regime ($ Q\ll1 $). In the weak dissipative regime the extra friction added by $ \Gamma $ is not enough to modify the background inflation evolution, and it will resemble that of cold inflation. \\
Using equations \eqref{friedman}, \eqref{conservation3} and quasi stable equation \eqref{density4} friedmann equation reads 
 \begin{equation}
\label{HJ1}
\dot{\phi}=-\frac{2\lambda+1}{\lambda+2}\frac{2}{Q+1}H_{2}^{'},
\end{equation}
Now, inserting this equation into the equation \eqref{friedman}, we can express the potential in the terms of the scalar field as follows
 \begin{equation}
\label{HJ2}
U(\phi)=(1+2\lambda)H_{2}^{2}-2\left(\frac{2\lambda +1}{\lambda +2}\frac{1}{(Q+1)}H_{2}^{'}\right)^{2}.
\end{equation}
Equation \eqref{HJ1} and \eqref{HJ2} are well known Hamilton-Jacobi formalism.  The slow-roll conditions are given by \cite{Bastero-Gil1}
 \begin{equation}
\label{slowroll1}
\varepsilon=\frac{6(2\lambda+1)}{(\lambda+2)^{2}}\left( \frac{H_{2}^{'}}{H_{2}}\right) ^{2}\frac{1}{Q+1}
\end{equation}
 \begin{equation}
\label{slowroll2}
\eta=\frac{12(2\lambda+1)}{(\lambda+2)^{2}}\frac{H_{2}^{''}}{H_{2}}\frac{1}{Q+1}
\end{equation}
 \begin{equation}
\label{slowroll3}
\beta=\frac{6(2\lambda+1)}{(\lambda+2)^{2}}\frac{H_{2}^{'}}{H_{2}}\frac{\Gamma^{'}}{\Gamma}\frac{1}{Q+1}
\end{equation}
\begin{equation}
\label{slowroll4}
\sigma=\frac{6(2\lambda+1)}{(\lambda+2)^{2}}\frac{H_{2}^{'''}}{H_{2}}\frac{1}{Q+1}
\end{equation}
\begin{equation}
\label{slowroll5}
\delta=\left( \frac{6(2\lambda+1)}{(\lambda+2)^{2}}\frac{H_{2}^{'}}{H_{2}}\frac{1}{Q+1}\right)^{2}\frac{\Gamma^{'}}{\Gamma}.
\end{equation}
Knowing that at the end of inflation $ \varepsilon\simeq1 $, equation \eqref{slowroll1} can be used to find inflaton at the end of inflation $ \phi_{e} $. The fluid energy density can be calculated as follows 
\begin{equation}
\label{density}
\rho=\frac{1}{2}\frac{Q}{Q+1}\varepsilon \rho_{\phi}.
\end{equation}
The number of efolds is defined as 
\begin{equation}
\label{efolds}
N=-\frac{(\lambda+2)^{2}}{6(2\lambda+1)}\int_{\phi_{*}}^{\phi_{e}} \frac{H_{2}}{H_{2}^{'}}(Q+1)d\phi.
\end{equation}
By considering $ N\approx 50-60 $ e-fold to comfortably solve the problems of the standard big-bang cosmology, this equation can be used to find inflaton at the horizon crossing $ \phi_{*} $.  the scalar power spectrum is derived as 
\begin{equation}
\label{powerspec}
\mathcal{P}_{s}=\frac{25}{36}\frac{H_2^2(\lambda+2)}{\dot{\phi}^2}\delta\phi^{2}.
\end{equation}
and the scalar spectral index and the running of scalar spectral index is given by 
\begin{equation}
\label{scalarspectralindex}
n_{s}-1=\frac{d \ln \mathcal{P}_{s}}{d\ln k}
\end{equation}
  \begin{equation}
\label{Running}
\alpha_{s}=\frac{d \ln n_{s}}{d\ln k}
\end{equation}
The generation of tensor perturbation give rise to stimulated emission in the thermal background of gravitational waves during inflation. This process changes the tensor power spectrum by an extra temperature dependent factor $ \coth [k/2T ] $ \cite{Bhattacharya}. Thus it is given by
   \begin{equation}
\label{tensor}
 \mathcal{P}_{t}=2\left( \frac{(\lambda+2)H_{2}}{6\pi}\right) ^{2}\coth [\frac{k}{2T}],
\end{equation}
the tensor spectral index being
  \begin{equation}
\label{tensorspectral index}
n_{t}=\frac{d}{d\ln k}\ln \left[ \frac{\mathcal{P}_{t}}{\coth [k/2T]}\right].
\end{equation}
Moreover the quantity of interest is the tensor scalar ratio given by
  \begin{equation}
\label{TSratio}
r(k_{*})=\left( \frac{\mathcal{P}_{t}}{\mathcal{P}_{s}}\right)|_{k=k_{*}}
\end{equation}
where $ k_{*}=0.05 Mpc^{-1} $ is the wave number at the time when the pivot scale exits the Hubble radius. We restrict ourselves to study the high dissipative regime ($ Q\gg1 $). In this case the coefficient ($ 1+Q $) could be approximated as $ Q $. In warm inflation, thermal interaction with the matter-radiation field produce fluctuations of the scalar field, therefore we have
  \begin{equation}
\label{phi}
\delta\phi^{2}= \frac{k_{F}T}{2\pi^{2}}, 
\end{equation}
where $ k_{F}=\sqrt{\Gamma(\lambda+2)H_{2}/3}=\sqrt{Q/3}(\lambda+2)H_{2}\geq H_{2} $ is the freeze-out scale at which dissipation damps out the thermally excited fluctuations \cite{Taylor}. Substituting \eqref{phi} in Eq. \eqref{powerspec}, we get the scalar power spectrum in strong dissipative regime as
\begin{equation}
\label{powerspec2}
\mathcal{P}_{s}=\frac{25}{856\pi^2}\left( \frac{(\lambda+2)^3\Gamma^3H_2}{(2\lambda+1)^2C_{\gamma}^{1/3}H_2^{'2}}\right) ^{3/4}.
\end{equation}
 Then, scalar spectral index and its running could easily be derived as
  \begin{equation}
\label{scalar22}
n_{s}=1-\frac{3}{4}\left( 3\tilde{\beta}-\tilde{\eta}+\tilde{\varepsilon}\right), 
\end{equation}  
  \begin{equation}
  \begin{split}
\label{running22}
\alpha_{s}=\frac{3}{4}\left( \tilde{\eta}\tilde{\varepsilon}-\tilde{\varepsilon}^{2}-2\tilde{\varepsilon}\tilde{\sigma}
-\tilde{\varepsilon}\tilde{\beta}+\frac{5}{2}\tilde{\beta}\tilde{\eta}+3\tilde{\delta}-6\tilde{\beta}^{2}\right) 
,
\end{split}
\end{equation}
where the tilde represent the high dissipative regime. Moreover, the tensor-to-scalar ratio is evaluted as
\begin{equation}
\label{tsratio1}
r=\frac{88}{75}\left(\frac{(\lambda+2)^{1/3}H_2^{5/3}C_{\gamma}^{1/3}H_2^{'2}}{(2\lambda+1)\Gamma^3} \right)^{3/4}\coth\left[ \frac{k}{2T}\right]  .
\end{equation}
\subsection{Typical example for warm anisotropic inflation}
\label{S:3:2}
\begin{figure}
 \includegraphics[width=.5\linewidth]{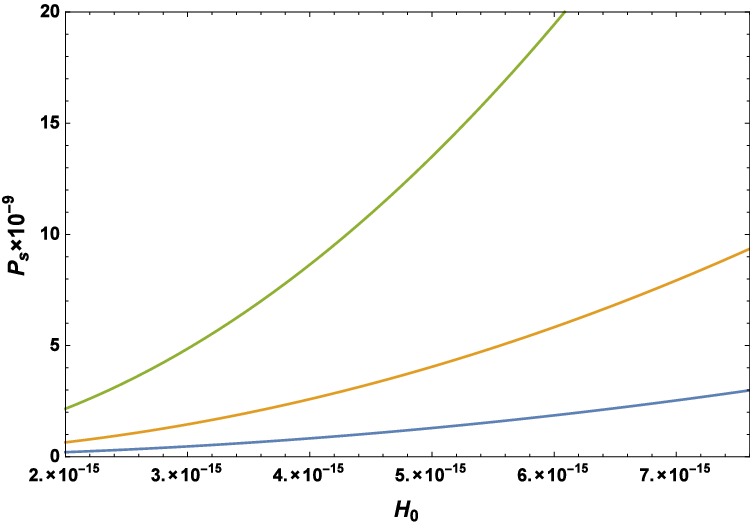} \centering
   \caption{$ \mathcal{P}_{s} $ as a function of $ H_0 $ for $ \lambda=1 $ (green), $ \lambda=1.5 $ (orange) $ \lambda=2 $ (blue). The other constant parameters are $ n=0.5 $, $ m=3 $, $ N=55 $, $ \theta=2^{10} $, and $ k=0.05 $. }
   \label{H0Gamma0withoutViscusity}
\end{figure}

\begin{figure}[h]
 \centering
 \subfigure[$r-n_s$ diagram]{\includegraphics[width=6cm]{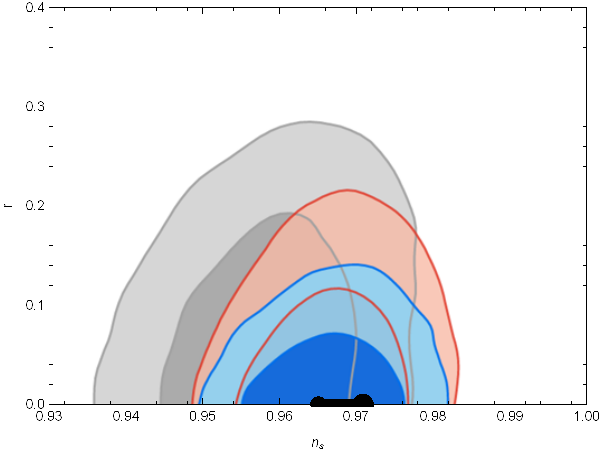}}
 \subfigure[$\alpha_s-n_s$ diagram]{\includegraphics[width=6cm]{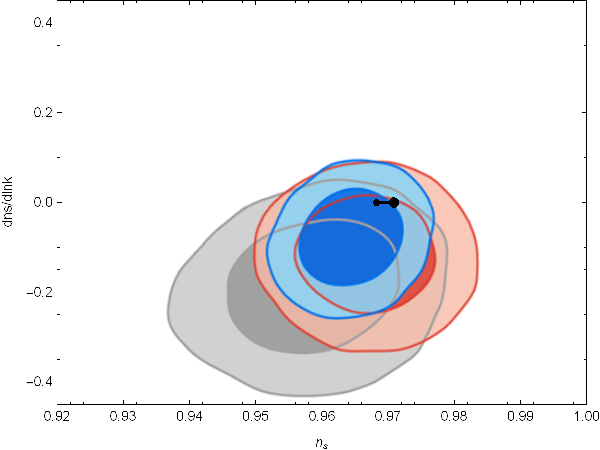}}
   \caption{$r-n_s$ and b)$\alpha_s-n_s$ diagrams of the model is compared with Planck data  for $ \lambda=1.5 $. The small point belongs to $N=50$ and the large point corresponds to $N=60$. The other constant parameters are $ n=0.5 $, $ m=3 $, $ \theta=2^{10} $, $ k=0.05 $.  }
   \label{rnsm3n05lambda2withoutViscousity}
\end{figure}

\begin{figure}
 \includegraphics[width=.5\linewidth]{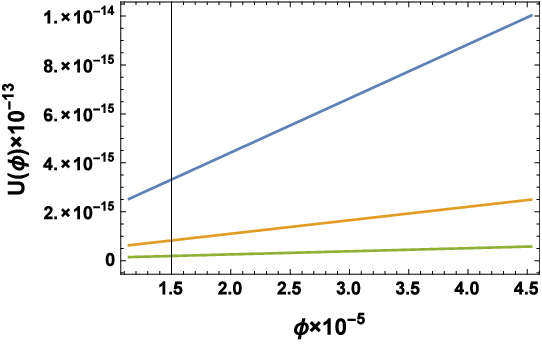} \centering
   \caption{The potential as a function of scalar field during the inflationary time  for $ \lambda=1 $ (green), $ \lambda=1.5 $ (orange) and $ \lambda=2 $ (blue), where the constant parameters are $ \gamma=1.5 $, $ n=0.5 $, $ m=3 $, $ N=55 $, $ \theta=2^{10} $, , $ k=0.05 $.  }
   \label{Um3n05lamdadiffwithoutViscous}
\end{figure}

\begin{figure}
 \includegraphics[width=.5\linewidth]{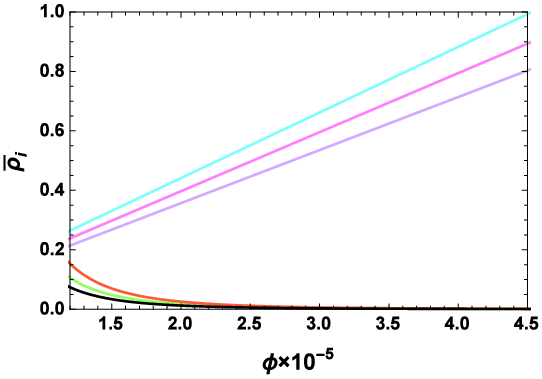} \centering
   \caption{The energy density of the scalar field as a function of scalar field  for $ \lambda=1 $ (purple), $ \lambda=1.5 $ (pink) and $ \lambda=2 $ (blue) and the energy density of the fluid for $ \lambda=1 $ (black), $ \lambda=1.5 $ (green) and $ \lambda=2 $ (red). The other initial value are the same as Fig. \ref{Um3n05lamdadiffwithoutViscous}.   }
   \label{densityform3n05lambdadiffwithoutViscous}
\end{figure}

 In this section, we will consider a typical example. We would consider Hubble parameter in a power-low function of the scalar field, namley 
\begin{equation}\label{Hubble}
H(\phi)=H_0\phi^n.
\end{equation}
By this consideration the potential will be power-law too. Power-low potential is of great importance because it is in good agreement with observational data. We will consider a specific case for $ \Gamma $, and (as mentioned above) we restrict ourselves to study the strong dissipative regime ($ Q\gg1 $). The parameters of the model will be restricted by comparison to the observational data.
When the dissipation coefficient is power-low $ \Gamma=\Gamma_{0}\phi^m $, the scalar field at the horizon crossing and end of inflation, will be obtained as 
\begin{equation}\label{phiend}
 \phi_*=\phi_e\left(\frac{N(2-n+m)}{n}+1 \right)^{\frac{1}{2-n+m}}, \  \ \phi_e=\left( \frac{6(2\lambda+1)n^2 H_0}{(\lambda+2)\Gamma_{0}} \right)^{\frac{1}{2-n+m}}.
\end{equation}
The slow-roll parameters at the horizon crossing read
\begin{equation*}\label{slowroll22}
\varepsilon^*=\frac{n}{(2-n+m)N+n},\  \ \eta^*=2\left(\frac{n-1}{n} \right)\varepsilon^*, \  \ \sigma^*=\frac{1}{2}\left(\frac{n-2}{n} \right)\eta^*,
\end{equation*}
\begin{equation}\label{slowroll222}
\beta^*=\frac{m}{n}\varepsilon^*, \  \ \delta^*=\frac{m(m-1)}{n^2}\varepsilon^{*2}.
\end{equation}
In the strong dissipative regime ($ Q\gg 1 $) we have $ \frac{\Gamma_0}{H_0}\gg(\lambda+2)\phi^{n-m} $. This can be simplified as
\begin{equation}\label{s}
 \frac{\Gamma_0}{H_0}= \theta^{\nu/2}(\lambda+2)^{\nu/2}\left( \frac{6n^2 (2\lambda+1)}{\lambda+2}(\frac{\nu}{n}N+1)\right)^{(-\nu+2)/2} 
\end{equation}
where $ \theta $ is a constant and it is chosen to be much bigger than  one to satisfy $ Q\gg 1 $, and $ \nu=m-n+2 $. In order to obtain $ H_0 $, using \eqref{s} we plot $ \mathcal{P}_{s} $ as a function of $ H_0 $. It has been shown in the Fig. \ref{H0Gamma0withoutViscusity} for $ n=0.5 $. This diagram shows that $ H_0=2.06\times10^{-15} $, $ 3.71\times10^{-15} $ and $ 6.65\times10^{-15} $ respectively  for $ \lambda=1 $, $ 1.5 $ and $ 2 $. Using \eqref{s}, $ \Gamma_0 $ is obtained as $ 2.075\times10^{7} $, $ 6.903\times10^{7} $ and $2.175\times10^{8} $ respectively.
Using this initial values spectral index and its running will be obtaied as $ n_s=0.968246 $ and $ \alpha=-0.000558 $ for $ N=55 $. It has been shown in the Fig. \ref{rnsm3n05lambda2withoutViscousity}. This figure shows that $ r-n_s $ is in the acceptable rang.
 Standing $ n_s $ in the acceptable range is not enogh to choose the model as a suitable model to describe the warm inflationary scenario. The potential energy and the ratio of energy density should be examined too. They have been shown in Fig. \ref{Um3n05lamdadiffwithoutViscous} , \ref{densityform3n05lambdadiffwithoutViscous}. From Fig. \ref{Um3n05lamdadiffwithoutViscous} it is clear that the potential stands below the Plank energy scale in all cases for $ \lambda $. Further, the scalar field falls down to the minimum of the potential.  Fig. \ref{densityform3n05lambdadiffwithoutViscous} shows that at the horizon crossing, energy density of the scalar field is bigger then that of the fluid. However, at the end of inflation, energy density of the scalar field and the fluid come closer to each other and the radiation energy density increases with decreasing of the energy density of the scalar field. This point matches to the condition of warm inflation. 
 

\section{Warm anisotropic inflationary model with viscous pressure}
\label{S:4}

\subsection{Generality}
Enegy density and pressure of the inflaton are the same as \eqref{density11}.
But total pressure of the imperfect fluid is changed as follows
\begin{equation}
\label{density}
\rho=Ts(\phi, T), \  \ p=(\gamma-1)\rho+\Pi,
\end{equation}
where $ T $ stands for temperature and $ s=s(\phi, T) $ for entropy density of imperfect fluid, and it is defined by a thermodynamical relation 
\begin{equation}
\label{entropy}
s(\phi, T)=-\frac{\partial f}{\partial T}=-\frac{\partial U(\phi, T)}{\partial T}, \  \ f=\rho_{\phi}+\rho-Ts
\end{equation}
where $ f $ is the Helmholtz free energy which is dominated by $ U(\phi, T) $ during slow-roll approximation. The subindex $ r $ has been omited.
For the viscous pressure $ \Pi $, we shall assume the usual fluid dynamics expression $ \Pi=-(\lambda+2)\zeta H_{2} $ \cite{Huang}, where $ \zeta $ denotes the phenamenological coefficient of bulk viscosity. According to second law of thermodynamics, this coefficient is positive-definite quantity and in general it is expected to depend on the energy density of the fluid. 
\\
The corresponding Friedmann equation is the same as \eqref{friedman}.
But the conservation equations for the inflaton and the imperfect fluid reads
\begin{equation}
\label{conservation1}
\dot{\rho}_{\phi}+(\lambda+2)H_{2}(\rho_{\phi}+P_{\phi})=-\Gamma(\rho_{\phi}+P_{\phi}) ,
\end{equation} 
\begin{equation}
\label{conservation2}
\dot{\rho}+(\lambda+2)H_{2}(\gamma \rho+\Pi)=\Gamma(\rho_{\phi}+P_{\phi}).
\end{equation}
where $ \gamma=4/3 $, which is valid for a quasi-equilibrium high temperature thermal bath typical of warm inflation. 
 Eq. \eqref{conservation1} is equivalent to the evolution equation for the inflaton field in \eqref{conservation3}.
Using equation \eqref{conservation2}, and that the entropy density $ s $ is related to the energy density by $ Ts=\gamma\rho $, we can find the entropy density evolution for the model as follows
\begin{equation}
\label{conservationentropy}
T\dot{s}+(\lambda+2)H_{2}(Ts+\Pi)=\Gamma \dot{\phi}^{2}.
\end{equation}
This equation reads the dynamical effects of the bulk viscosity. If the bulk viscous pressure $ \Pi $ is negative, it decreases the radiation pressure, thus enhancing the effect from the source term on the RHS of \eqref{conservationentropy}. As a consequence, the entropy density increases, and therefore the radiation energy density grows. on the one hand, if this bulk pressure term is too large, there is too much radiation production and the radiation energy density dominates too soon over the scalar field energy density, thus spoiling inflation. This regime is called the unstable regime. On the other hand, if the bulk pressure term is controlled to avoid the radiation domination until the end of inflation, the system is said to be in the stable regime. In this regime the bulk viscosity give rise to an additional negative pressure, and hence, inflation is enhanced. So the stable regime is obtained by using three limits. First by using $ \rho_{\phi}\thickapprox U(\phi) $ and $ \rho_{\phi}>\rho $, second using slow-roll approximation $ \dot{\phi}^{2} \ll U(\phi) $, $ \ddot{\phi}^{2} \ll ((\lambda +2)H_{2}+\gamma )\dot{\phi} $ and third by considering quasi-stable condition \cite{Bastero-Gil}. Using this limits \eqref{friedman}, \eqref{conservation2} and \eqref{conservation3}  are reduced to
 \begin{equation}
\label{friedman2}
H_{2}^{2}=(\frac{1}{\lambda+2})U(\phi),
\end{equation}
 \begin{equation}
\label{conservation33}
H_{2}\dot{\phi}=-\frac{1}{\lambda+2}\frac{U^{'}(\phi)}{Q+1},
\end{equation}
 \begin{equation}
\label{conservation22}
\gamma\rho-\Pi=Q\dot{\phi}^2,
\end{equation}
The Hamilton-Jacobi equation is obtained the same as \eqref{HJ1} and \eqref{HJ2}. Moreover the slow-roll conditions are the same as \eqref{slowroll1}-\eqref{slowroll5}. 
Equation \eqref{HJ1} and \eqref{slowroll1} in equation \eqref{conservation22}, the fluid energy density can be calculated as
\begin{equation}
\label{density}
\rho =Ts=\frac{1}{\gamma}\left( \frac{2}{3}\frac{Q}{Q+1}\varepsilon \rho_{\phi}-\Pi \right) .
\end{equation}
The warm anisotropic inflation with viscous pressure could take place when $ \varepsilon<1 $ so 
\begin{equation}
\label{density1}
\rho _{\phi}>\frac{3}{2}\frac{Q+1}{Q}( \gamma\rho+\Pi).
\end{equation}
Inflation ends when $ \varepsilon\simeq1 $ which reads 
\begin{equation}
\label{density2}
\rho _{\phi}=\frac{3}{2}\frac{Q+1}{Q}( \gamma\rho+\Pi).
\end{equation}
The number of efolds is the same as \eqref{efolds}, too. But the scalar power spectrum under slow-roll condition is as fallows 
\begin{equation}
\label{Spower}
\mathcal{P}_{s}=\frac{64\pi^{2}e^{-2\chi(\phi)}}{U^{'}(\phi)^{2}}\delta\phi^{2}.
\end{equation}
 where the auxiliary function $ \chi(\phi) $ is
\begin{equation}
\begin{aligned}
\chi(\phi) =&-\int \left( \frac{\Gamma^{'}}{\Gamma}\frac{Q}{Q+1}+\frac{U^{'}(\phi)3}{U(\phi)8G(\phi)}\frac{Q+2}{(Q+1)^{2}}\left[(Q+\frac{4}{3})-  \right. \right.\\
&\left. \left. \left(\gamma-1+\Pi\frac{\xi_{,\rho}}{\xi} \right)\frac{\Gamma^{'}U^{'}(\phi)}{\gamma(Q+1)H_2^{3} (\lambda+2)^3}\right]\right)d\phi.
\end{aligned}
\label{auxiliary}
\end{equation}
 and 
  \begin{equation}
\label{auxiliary2}
G(\phi)=1-\frac{9}{8H_{2}^{2}(\lambda+2)}\left(2\gamma\rho+3\Pi+\frac{\gamma\rho+\Pi}{\gamma}\left[\frac{\xi_{,\rho}}{\xi}\Pi-1 \right]  \right) 
\end{equation}

The scalar spectral index $ n_{s} $ is given by
 \begin{equation}
\label{scalarspectralindex}
n_{s}-1=\frac{d \ln \mathcal{P}_{s}}{d\ln k}
\end{equation}
 Running of the scalar spectral index is another cosmological parameter which is obtained by taking derivative of the scalar spectral index as follows
  \begin{equation}
\label{Running}
\alpha_{s}=\frac{d \ln n_{s}}{d\ln k}
\end{equation}

 Substituting \eqref{phi} in Eq. \eqref{Spower}, scalar perturbation for strong dissipative regime reads
   \begin{equation}
\label{Spower2}
\mathcal{P}_{s}= \frac{8}{(2\lambda+1)^{2}}\frac{((\lambda+2)\Gamma/3)^{1/2}T_{r}}{H^{'2}_{2}H_{2}^{3/2}}\; e^{-2\tilde{\chi}(\phi)}, 
\end{equation} 
 where $ \tilde{\chi}(\phi):=\chi(\phi)|_{Q\gg1} $ is as follows
    \begin{equation}
\label{Spower2}
\tilde{\chi}(\phi)=-\int \left( \frac{\Gamma^{'}}{\Gamma}+\frac{3U^{'}}{8UG(\phi)}\left[ 1-\left(\gamma-1+\Pi\frac{\xi_{,\rho}}{\xi} \right)\frac{\Gamma^{'}U^{'}}{\gamma\Gamma^{2}(\lambda+2)H_2} \right] \right) d\phi. 
\end{equation} 
 Using \eqref{scalarspectralindex}, \eqref{Running} and \eqref{Spower2} scalar spectral index and its running could be derived as
 \begin{equation}
\label{scalar2}
n_{s}=1-\frac{1}{2}\tilde{\beta}+\tilde{\eta}+\frac{3}{2}\tilde{\varepsilon}+12\frac{2\lambda+1}{(\lambda+2)^{2}}\frac{H_{2}^{'}}{QH_{2}}\tilde{\chi}^{'}(\phi)
\end{equation}  
 and
  \begin{equation}
  \begin{split}
\label{running2}
\alpha_{s}=\frac{3}{2}(-\tilde{\eta}\tilde{\varepsilon}+\tilde{\varepsilon}^{2}+\tilde{\varepsilon}\tilde{\beta})-2\tilde{\sigma}\tilde{\eta}+\frac{5}{4}\tilde{\beta}\tilde{\eta}+\frac{1}{2}\tilde{\delta}-\tilde{\beta}^{2}&\\+(2\tilde{\beta}-\tilde{\eta})\tilde{\varepsilon}\frac{H_{2}}{H_{2}^{'}}\tilde{\chi}^{'}(\phi)-12\frac{2\lambda+1}{(\lambda+2)^{2}}\frac{\varepsilon}{Q}\tilde{\chi}^{''}(\phi)
\end{split}
\end{equation}

The tensor power spectrum, the tensor spectral index and the tensor-to-scalar ratio are the same as \eqref{tensor},  \eqref{tensorspectral index} and \eqref{TSratio}.
Substituting \eqref{tensor}, \eqref{scalar2} in \eqref{TSratio} the tensor-to-scalar ratio is evaluted as 
\begin{equation}\label{stratio2}
r (k_{*})={(\lambda+2)^2 \over 16\pi^2 }{\sqrt{3}(2\lambda+1)^2 \over (\lambda+2)^{1/2}} { H^{7/2} H'^2 \over \Gamma^{1/2} T_r } \; \exp[2\tilde\chi(\phi)] \coth\Big({k \over 2T}\Big)|_{k=k_{*}}.
\end{equation}


\subsection{Typical example for warm anisotropic viscous inflation}
\label{S:4:2}
\begin{figure}
 \includegraphics[width=.5\linewidth]{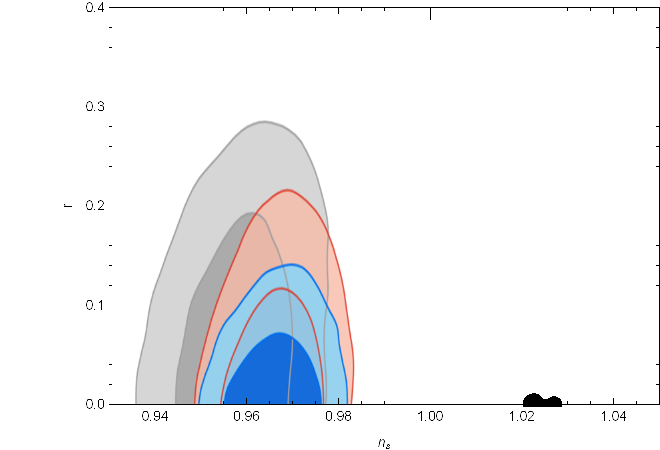} \centering
   \caption{. }
   \label{H0Gamma0n1landa5}
\end{figure}

\begin{figure}
 \includegraphics[width=.5\linewidth]{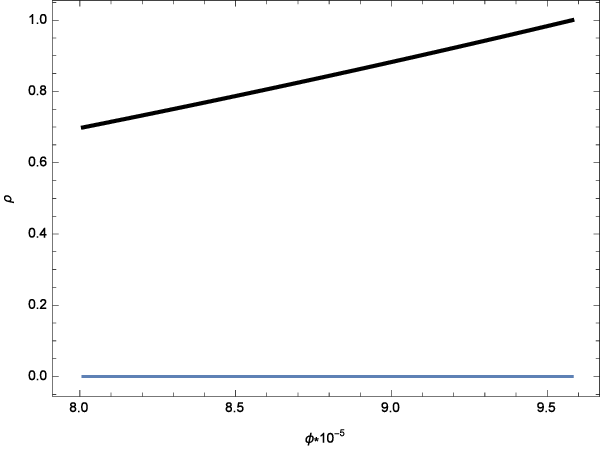} \centering
   \caption{ }
   \label{rho}
\end{figure}

As the previeus section the Hubble parameter is considered to be power-low function of the scalar field. Also, we will consider the strong dissipative regime. Moreover we will consider
 $ \Gamma=\Gamma_{0}\phi^m $ and $ \xi=\xi_{0} $.
Let us first consider both dissipative coefficient and bulk viscous coefficient to be constant i.e. $ m=0 $. Imposing $ \varepsilon\simeq1 $ provides the inflaton at the end of inflation as
\begin{equation}\label{phiend}
\phi_e=\left( \frac{6(2\lambda+1)n^2 H_0}{(\lambda+2)\Gamma_{0}} \right)^{\frac{1}{2-n}}.
\end{equation}
From Eq. \eqref{efolds} inflaton at the horizon crossing will be obtained as
\begin{equation}
\label{phicrossing}
\phi_*=\phi_e\left(\frac{N(2-n)}{n}+1 \right)^{\frac{1}{2-n}} 
\end{equation}
The non-vanishing slow-roll parameters at horizon crossing read
\begin{equation*}
\varepsilon^*=\frac{n}{(2-n)N+1},\  \ \eta^*=2\left(\frac{n-1}{n} \right)\varepsilon^*, \  \ \sigma^*=\frac{1}{2}\left(\frac{n-2}{n} \right)\eta^*.
\label{slowroll11}
\end{equation*}
For the other two parameters we have: $ \beta^*=\delta^*=0 $. In order to obtain $ n_s $, $ \alpha_s $ and $ r $ we need to obtain $ H_0 $ and $ \Gamma_0 $. We fix the scalar power spectrum in Eq. \eqref{Spower2} and the tensor-to-scalar ratio in Eq. \eqref{TSratio} as $ \mathcal{P}^*_{s}=2.207\times 10^{-9} $ and $ r^*<0.11 $ respectivly. In this way, we find two equations that give values for the parameter $ H_0 $ and $ \Gamma_0 $ for $ n=1 $  as follows
\begin{equation*}
H_0^2=Z\Gamma_0
\label{H0}
\end{equation*}
\begin{equation*}
\Gamma_0=\left(\frac{8(\lambda+2)^{1/2}}{\sqrt{3}(2\lambda+1)^2} \frac{T\left(6(N+1)-\frac{1+2\lambda}{\lambda+2} \frac{9}{2}\frac{2\gamma-1}{\gamma}+\frac{9(\lambda+2)\xi_0}{8(2\lambda+1)Z} \right)^{3/2} }{(6(N+1))^{3/2}Z^{5/2}(6(2\lambda+1)/(\lambda+2))^{3/2}\mathcal{P}^*_{s}} \right) ^{4/5}
\label{H0}
\end{equation*}
where $ Z $ is defined as
\begin{equation}
\label{phicrossing}
Z= \left(\frac{\pi}{(\lambda+2)(N+1)} \right) \left(\frac{\mathcal{P}^*_{s}r^*}{2\coth \left[\frac{k}{2T} \right] } \right) ^{1/2}
\end{equation}
In this case for $ \lambda=1.5 $ the line stands far out of the acceptable range. It is shown in the Fig.\ref{H0Gamma0n1landa5}. Increasing $ \lambda $ give a worse result. However, in this case the energy density of the fluid is smaller then the energy density of the scalar field, Fig. \ref{rho} . It is consistant with the first assumption of the warm inflation scenario.\\

We now consider $ m\neq0 $. Using Eq. \eqref{slowroll1} and \eqref{efolds}, the scalar field at the horizon crossing and end of inflation are obtained the same as Eq. \eqref{phiend}.
The slow-roll parameters at the horizon crossing are obtained the same as Eq. \eqref{slowroll222}.

\begin{figure}[h]
 \centering
 \subfigure[$r-n_s$ diagram]{\includegraphics[width=6cm]{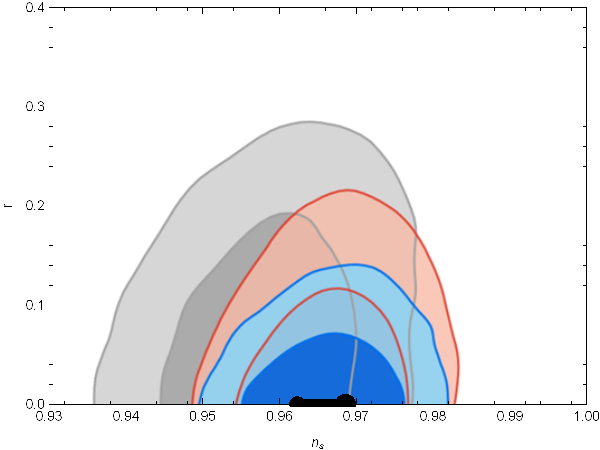}}
 \subfigure[$\alpha_s-n_s$ diagram]{\includegraphics[width=6cm]{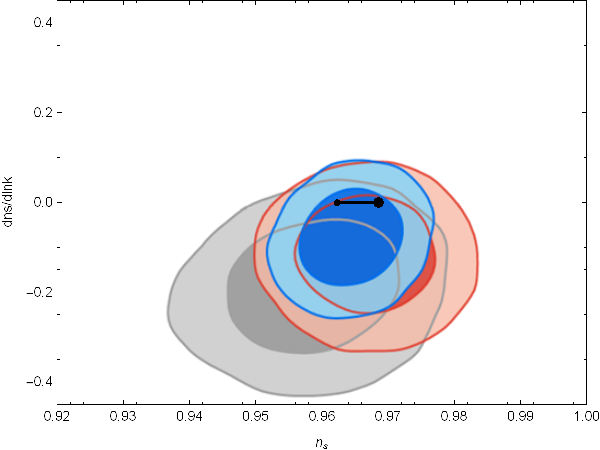}}
 \caption{a) $r-n_s$ and b)$\alpha_s-n_s$ diagrams of the model is compared with Planck data  for $ \lambda=2 $. The small point belongs to $N=50$ and the large point corresponds to $N=60$. The other constant parameters are $ \gamma=1.5 $, $ n=0.5 $, $ m=3 $, $ \theta=2^{10} $, $ \xi_0=7\times 10^{-14} $, $ k=0.05 $ and $ T_r=T=2\times 10^{-5} $.  }\label{rnsm3n05lambda2}
\end{figure}

\begin{figure}
 \includegraphics[width=.5\linewidth]{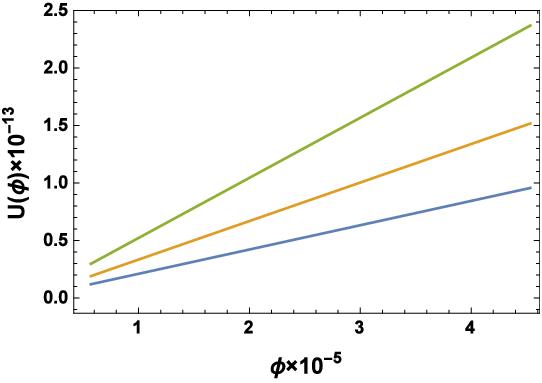} \centering
   \caption{The potential as a function of scalar field during the inflationary time  for $ \lambda=1 $ (green), $ \lambda=1.5 $ (orange) and $ \lambda=2 $ (blue), where the constant parameters are $ \gamma=1.5 $, $ n=0.5 $, $ m=3 $, $ N=55 $, $ \theta=2^{10} $, $ \xi_0=7\times 10^{-14} $, $ k=0.05 $ and $ T_r=T=2\times 10^{-5} $.  }
   \label{Um3n05lamdadiff}
\end{figure}

\begin{figure}
 \includegraphics[width=.5\linewidth]{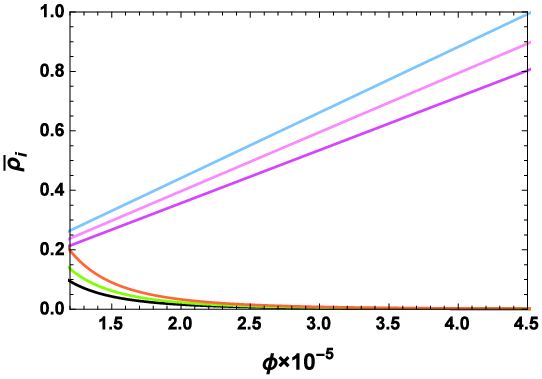} \centering
   \caption{The energy density of the scalar field as a function of scalar field  for $ \lambda=1 $ (purple), $ \lambda=1.5 $ (pink) and $ \lambda=2 $ (blue) and the energy density of the fluid for $ \lambda=1 $ (black), $ \lambda=1.5 $ (green) and $ \lambda=2 $ (red). The other initial value are the same as Fig. \ref{Um3n05lamdadiff}.   }
   \label{densityform3n05lambdadiff}
\end{figure}

 This diagram shows that $ H_0=1.32\times10^{-7} $, $ 9.15\times10^{-8} $ and $ 6.50\times10^{-8} $ respectively  for $ \lambda=1 $, $ 1.5 $ and $ 2 $. Putting these values in \eqref{s} $ \Gamma_0 $ is obtained easily. We should notice that $ \lambda=1 $ corresponds to the isotropic case.
\begin{figure}

 \includegraphics[width=.5\linewidth]{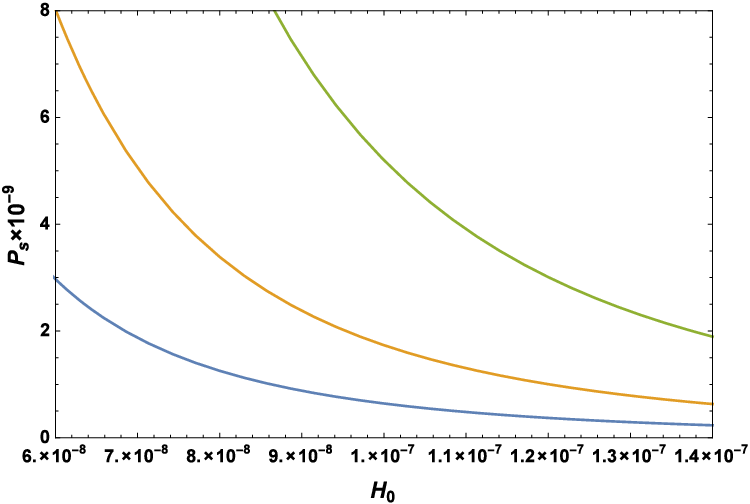} \centering
   \caption{$ \mathcal{P}_{s} $ as a function of $ H_0 $ for $ \lambda=1 $ (green), $ \lambda=1.5 $ (orange) $ \lambda=2 $ (blue). The other constant parameters are $ \gamma=1.5 $, $ n=0.5 $, $ m=3 $, $ N=55 $, $ \theta=2^{10} $, $ \xi_0=7\times 10^{-14} $, $ k=0.05 $ and $ T_r=T=2\times 10^{-5} $.  }
   \label{psm3n1lambdadiff}
\end{figure}
We could obtain spectra index and its running for the model respectivly as $ n_s=0.965723 $ and $ \alpha_s=-0.000575 $ for $ \lambda=1.5 $ and $ 2 $. It has been shown in Fig. \ref{rnsm3n05lambda2} for  $ \lambda=2 $. For the isotropic case we could obtain $ n_s=0.965726 $ and $ \alpha_s=-0.000597 $ which is the same as anisotropic case.  The potential energy and the ratio of energy density has been plotted, too. They have been shown in Fig. \ref{Um3n05lamdadiff} , \ref{densityform3n05lambdadiff}. They are the same as Fig. \ref{Um3n05lamdadiffwithoutViscous} and \ref{densityform3n05lambdadiffwithoutViscous}. It shows that the viscous  pressure doesn't affect potential energy and the ratio of energy density significantly.


\begin{figure}
 \includegraphics[width=.5\linewidth]{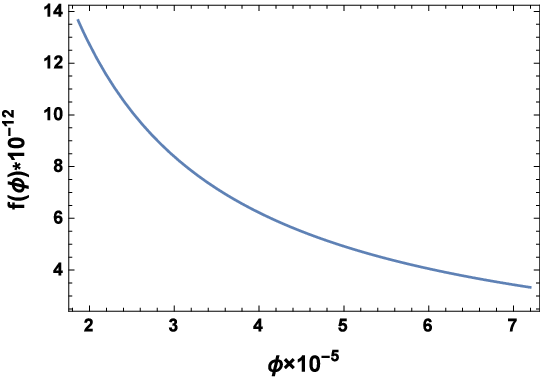} \centering
   \caption{The potential as a function of scalar field during the inflationary time  for   }
   \label{TemporHtest}
\end{figure}

Another important condition which should be satisfied during warm inflation is $ T_r >(\lambda+2)H_2/3 $. Using equation \eqref{radiationdensity}, this condition gives
\begin{equation}\label{THtest}
 f(\phi)= \left( \frac{4H_0 \phi^{n-2}}{\theta}-(\lambda+2)\xi-(\frac{\lambda+2}{3})^4 H_0^2\phi^{3n}\gamma C_{\gamma}\right)   >0
\end{equation}
This function has been shown in Fig. \ref{TemporHtest}, which shows that the condition $ T_r >(\lambda+2)H_2/3 $ is satisfied for all values of $ \phi $.

\section{Discussion and Conclusions}\label{conclusion}
In this work we have focuced on warm inflation within the anisotropic BI universe in the Hamilton-Jacobi formalism. We consider two different cases, with and without viscous pressure in the matter-radiation fluid. Viscous pressure arises either as a hydrdynamic effect or as a consequence of the decay of massive fields into light fields. A general relation for density perturbation, tensor perturbation, scalar spectral index, its running and the tensor-scalar ratio are obtaied in strong dissipative regime for both cases. In the case where viscous pressure is not considered an example was used to test the observational parameter. The Hubble parameter and the dissipation coefficient are given as a function of the scalar field. A suitable choice of the model parameters resulted in acceptable values for the scalar spectral index and its running for different choices of the anisotropic parameter $ \lambda $. $ r-n_s $ and $ dn_s/dlnk-n_s $ are plotted for a specific choices of the parameters of the model. It shows that  the model prediction for the perturbation parameters stands in the $ 68\%$   CL area, which predicts perfect agreement with Planck data. Moreover the anisotropic parameter $ \lambda $ doesn't have any significant effect on the purturbation parameters. The potential is plotted as function of the scalar field for different valuse of $ \lambda $. It shows that the potential stands below the Plank energy scale. Moreover, the dominance of the scalar field energy density over the radiation was examined and it shows that the constraint on the free paramters has been selected properly. Additionaly, for the selected parameters the energy densities of the scalar field and the fluid is ilustrated. They show that their behavior is compatible with warm inflation assumption.\\
In the next case warm inflation with viscous pressure is considerd. The Hubble parameter and dissipation coefficient are consided as before power low functions, But Bulk viscous coefficient is taken as constants. For sutible choices of the model parameters, purturbation parameters are obtained and  $ r-n_s $ and $ dn_s/dlnk-n_s $ are plotted. As the previous case the model prediction for the perturbation parameters stands in the $ 68\%$ CL area. How ever, it shows that by changing the anisotropic parameter $ \lambda $ no sighnificant effect occure for the perturbation parameters. The potential is plotted as function of the scalar field, too. It stands below the Plank energy scale. Moreover the energy densities of the scalar field and the fluid are cheked and shows good agreement with the warm inflation assumptions, too.










\end{document}